\documentclass[letterpaper, 10 pt, conference]{ieeeconf}
\IEEEoverridecommandlockouts

\usepackage{amsfonts,amsmath,amssymb} 

\DeclareMathOperator{\diag}{diag}         
\def\rbb{\mathbb{R}}
\def\trp{^T}
\def\diag{{\rm diag}}

\def\half{\frac{1}{2}}
\usepackage{psfrag,color}
\usepackage{enumerate,cite,latexsym,graphicx}
\newtheorem{theorem}{Theorem}

\newtheorem{definition}{Definition}

\title{\LARGE \bf A Reduced Order Direct Coupling Coherent Quantum Observer for a Complex Quantum Plant}

\author{Ian R.~Petersen and Elanor H. Huntington
\thanks{This work was supported by the
Australian Research Council (ARC) and the Chinese Academy of Sciences President’s International Fellowship Initiative (No. 2015DT006). }%
\thanks{Ian R. Petersen is with the School of  Engineering and Information Technology, 
        University of New South Wales at the Australian Defence Force Academy, Canberra ACT 2600, Australia.
         {\tt\small i.r.petersen@gmail.com} } 
\thanks{Elanor H. Huntington is with the 
College of Engineering and Computer Science, The Australian National University, Canberra, ACT 0200,
Australia. Email: Elanor.Huntington@anu.edu.au.}
}%

\begin{document}

\maketitle
\thispagestyle{empty}
\pagestyle{empty}

\begin{abstract}
This paper extends previous results on constructing a direct coupling quantum observer for a quantum harmonic oscillator system. In this case, we consider a complex linear quantum system plant consisting of a network of quantum harmonic oscillators. Conditions are given for which there exists a direct coupling observer which estimates a collection of variables in the quantum plant. It is shown that the order of the observer can be the same as the number of variables to be estimated when this number is even and thus this is a reduced order observer. 
\end{abstract}

\section{Introduction} \label{sec:intro}
A number of papers have recently considered the problem of constructing a coherent quantum observer for a quantum system; e.g., see \cite{MJ12a,VP9a,EMPUJ6a,PET14Aa}. In the coherent quantum observer problem, a quantum plant is coupled to a quantum observer which is also a quantum system. The quantum observer is constructed to be a physically realizable quantum system  so that the system variables of the quantum observer converge in some suitable sense to the system variables of the quantum plant. The papers \cite{PET14Aa,PET14Ba,PET14Ca,PET14Da}  considered the problem of constructing a direct coupling quantum observer for a given closed quantum system. In \cite{PET14Aa}, the proposed observer is shown to be able to estimate some but not all of the plant variables in a time averaged sense. Also, the paper \cite{PeHun1a} shows that a possible experimental implementation of the augmented quantum plant and quantum observer system considered in \cite{PET14Aa} may be constructed using a non-degenerate parametric amplifier (NDPA) which is coupled to a beamsplitter by suitable choice of the NDPA and beamsplitter parameters. 

In the paper \cite{PET14Aa}, the quantum plant consisted of a number of quantum harmonic oscillators where the number of variables to be estimated was allowed to be at most half of the total number of variables describing the quantum plant. However, the quantum plant was assumed to have very simple dynamics corresponding to a zero Hamiltonian. Then a quantum observer was constructed whose number of variables was equal to twice the number of variables to be estimated. In this paper we extend the results of \cite{PET14Aa} by first allowing for more general linear quantum plants with non-zero Hamiltonians. Conditions are given on whether a given set of variables of interest can be estimated via a direct coupling quantum observer. Then a direct coupling quantum observer is constructed whose order is the same as the number of variables to be estimated when this number is even. In the case that the number of variables to be estimated is odd, the order of the observer is one more than the number of variables to be estimated. Compared to the result in \cite{PET14Aa}, this is a reduced order observer. As in \cite{PET14Aa}, the convergence of the observer outputs to the plant outputs is a time averaged convergence since the overall plant-observer system is a closed quantum linear system. 

\section{Quantum  Systems}
In the  quantum observer problem under consideration, both the quantum plant and the  quantum observer are linear quantum systems; see also \cite{JNP1,GJ09,ZJ11}. We will restrict attention to closed linear quantum systems which do not interact with an external environment. 
The quantum mechanical behavior of a linear quantum system is described in terms of the system \emph{observables} which are self-adjoint operators on an underlying infinite dimensional complex Hilbert space $\mathfrak{H}$.   The commutator of two operators $x$ and $y$ on ${\mathfrak{H}}$ is  defined as $[x, y] = xy - yx$.~Also, for a  vector of operators $x$ on ${\mathfrak H}$, the commutator of ${x}$ and a scalar operator $y$ on ${\mathfrak{H}}$ is the  vector of operators $[{x},y] = {x} y - y {x}$, and the commutator of ${x}$ and its adjoint ${x}^\dagger$ is the  matrix of operators 
\[ [{x},{x}^\dagger] \triangleq {x} {x}^\dagger - ({x}^\# {x}^T)^T, \]
where ${x}^\# \triangleq (x_1^\ast\; x_2^\ast \;\cdots\; x_n^\ast )^T$ and $^\ast$ denotes the operator adjoint. 

The dynamics of the closed linear quantum systems under consideration are described by non-commutative differential equations of the form
\begin{eqnarray}
\dot x(t) &=& Ax(t); \quad x(0)=x_0
 \label{quantum_system}
\end{eqnarray}
where $A$ is a real matrix in $\rbb^{n
\times n}$, and $ x(t) = [\begin{array}{ccc} x_1(t) & \ldots &
x_n(t)
\end{array}]\trp$ is a vector of system observables; e.g., see \cite{JNP1}. Here $n$ is assumed to be an even number and $\frac{n}{2}$ is the number of modes in the quantum system. 

The initial system variables $x(0)=x_0$ 
are assumed to satisfy the {\em commutation relations}
\begin{equation}
[x_j(0), x_k(0) ] = 2 i \Theta_{jk}, \ \ j,k = 1, \ldots, n,
\label{x-ccr}
\end{equation}
where $\Theta$ is a real skew-symmetric matrix with components
$\Theta_{jk}$.  In the case of a
single quantum harmonic oscillator, we can choose $x=(x_1, x_2)^T$ where
$x_1=q$ is the position operator, and $x_2=p$ is the momentum
operator.  The
commutation relations are  $[q,p]=2 i$.
In general, the matrix $\Theta$ is assumed to be  of the  form
\begin{equation}
\label{Theta}
\Theta=\diag(J,J,\ldots,J)
\end{equation}
 where $J$ denotes the real skew-symmetric $2\times 2$ matrix
$$
J= \left[ \begin{array}{cc} 0 & 1 \\ -1 & 0
\end{array} \right].$$

The system dynamics (\ref{quantum_system}) are determined by the system Hamiltonian  
which is a self-adjoint operator on the underlying  Hilbert space  $\mathfrak{H}$. For the linear quantum systems under consideration, the system Hamiltonian will be a
quadratic form
$\mathcal{H}=\half x\trp R x$, where $R$ is a real symmetric matrix. Then, the corresponding matrix $A$ in 
(\ref{quantum_system}) is given by 
\begin{equation}
A=2\Theta R \label{eq_coef_cond_A}.
\end{equation}
 where $\Theta$ is defined as in (\ref{Theta}).
e.g., see \cite{JNP1}.
In this case, the  system variables $x(t)$ 
will satisfy the {\em commutation relations} at all times:
\begin{equation}
\label{CCR}
[x(t),x(t)^T]=2{\pmb i}\Theta \ \mbox{for all } t\geq 0.
\end{equation}
That is, the system will be \emph{physically realizable}; e.g., see \cite{JNP1}.

\section{Analysis of the Quantum Plant}
\label{sec:analysis}
In this section we will describe the class of quantum linear systems which will be considered as quantum plants. Also, we will analyse these quantum plants in order to provide conditions under which there exists a direct coupling observer which can estimate the quantum plant outputs. 

We consider general {\em closed linear quantum plants} described by
linear quantum system models of the following form:
\begin{eqnarray}
\dot x_p(t) &=& A_px_p(t); \quad x_p(0)=x_{0p}; \nonumber \\
z_p(t) &=& C_px_p(t)
 \label{plant}
\end{eqnarray}
where $z_p$ denotes the vector of system variables to be estimated by the observer and $ A_p \in \rbb^{n_p
\times n_p}$, $C_p\in \rbb^{m \times n_p}$. 
It is assumed that this quantum plant is physically realizable and corresponds to a plant Hamiltonian
$\mathcal{H}_p=\half x_p\trp R_p x_p$ where $R_p$  is a symmetric matrix and $A_p = 2 \Theta_p R_p$. Here $\Theta_p$ is of the form (\ref{Theta}). Unlike the case in \cite{PET14Aa}, we will not require that $R_p$ is zero. However, we will assume that $\det R_p = 0$ so that $R_p$ has a non-trivial null space. In addition, we assume that the matrices $R_p$ and $C_p$ satisfy the following conditions:
\begin{eqnarray}
\label{TF_cond}
C_p(sI-\Theta_p)^{-1}R_p &\equiv& 0;\\
\label{CpJnull}
C_p \Theta_p C_p^T &=& 0; \\
\label{Cprank}
\mbox{The matrix }C_p\mbox{ is of rank $m$.}
\end{eqnarray}
Note that if the matrix $C_p$ is not full rank then some of the components of $z_p$ can be expressed as linear combinations of the other components of $C_p$. Hence, without loss of generality, we can eliminate these components of $z_p$ to obtain a full rank $C_p$. 

In the sequel, we will show that these conditions imply that there exists a direct coupling quantum observer which can estimate the variables $z_p$. However, we first analyse the quantum plant satisfying these conditions. Indeed, we first consider the controllability of the pair $(\Theta_p, R_p)$. Since $\Theta_p^2 = -I$, it follows that the corresponding controllability matrix is given by 
\begin{eqnarray*}
\mathcal{C} &=& \left[\begin{array}{ccccc}R_p & \Theta_p R_p & \Theta_p^2 R_p &\ldots & \Theta_p^{n_p - 1}R_p\end{array}\right] \nonumber \\
&=& \left[\begin{array}{cccccc}R_p & \Theta_p R_p & - R_p  & -\Theta_p R_p&R_p &\ldots\end{array}\right];
\end{eqnarray*}
e.g., see \cite{RUG96}. 
This matrix has the same range space as the matrix 
\begin{equation}
\label{scriptCr}
\mathcal{C}_r = \left[\begin{array}{cc}R_p & \Theta_p R_p \end{array}\right].
\end{equation}
The range space of the matrix $\mathcal{C}_r$ will determine which variables of the quantum plant remain constant if the plant is not coupled to the quantum observer. These variables are ones which can be estimated by the quantum observer. 
 We can use the matrix $ \mathcal{C}_r$ to transform the pair $(\Theta_p, R_p)$ into a form corresponding to  controllable and uncontrollable 
subsystems; e.g., see \cite{RUG96}.  Indeed, we construct an orthogonal matrix $P$ using the svd of the matrix $ \mathcal{C}_r$ as $ \mathcal{C}_r = PSV^T$ where $V$ is also an orthogonal matrix and $S$ is a diagonal matrix. This construction of $P$ yields
\[
P^T\Theta_p P = \left[\begin{array}{cc} \Theta_{11} & \Theta_{12} \\ 0 & \Theta_{22} \end{array}\right], ~~ P^T R_p  = \left[\begin{array}{c} R_{p1} \\ 0  \end{array}\right]
\]
where the pair $(\Theta_{11},R_{p1})$ is controllable. 
Here $\Theta_{11} \in \rbb^{n_{p1} \times n_{p1}}$ and $\Theta_{22} \in \rbb^{n_{p2} \times n_{p2}}$ such that $ n_{p1}+ n_{p2} = n_p$. 

We now use the fact that $\Theta_p$ is a skew-symmetric matrix and hence $P^T\Theta_p P$ is a skew-symmetric matrix. Therefore, we must have
\begin{equation}
\label{transformed_theta}
P^T\Theta_p P = \left[\begin{array}{cc} \Theta_{11} & 0 \\ 0 & \Theta_{22} \end{array}\right]
\end{equation}
where $\Theta_{11}$ is skew-symmetric and $\Theta_{22}$ is skew-symmetric. Also, since the matrix $\Theta_p$ is non-singular, the matrices $\Theta_{11}$  and $\Theta_{22}$ must be non-singular. 

We also use the fact that $R_p$ is a symmetric matrix. To do this, we first  write $P = \left[\begin{array}{cc} P_{11} & P_{12} \\ P_{21} & P_{22} \end{array}\right]$ and
\[
P^T R_p = \left[\begin{array}{cc} R_{11} & R_{12} \\ 0 & 0 \end{array}\right]. 
\]
Hence,
\begin{eqnarray*}
P^T R_p P &=& \left[\begin{array}{cc} R_{11} & R_{12} \\ 0 & 0 \end{array}\right] \left[\begin{array}{cc} P_{11} & P_{12} \\ P_{21} & P_{22} \end{array}\right]\\
 &=& \left[\begin{array}{cc}R_{11} P_{11}+ R_{12}P_{21} & R_{11}P_{12}+R_{12}P_{22}\\0 & 0\end{array}\right].
\end{eqnarray*}
However, $R_p$ is symmetric and hence $P^T R_p P$ is symmetric. Thus, the matrix $P^T R_p P$ must be of the form
\begin{equation}
\label{Rpt}
P^T R_p P=\left[\begin{array}{cc}R_{p11}  & 0\\0 & 0\end{array}\right]
\end{equation}
where the matrix $R_{p11}$ is symmetric. Also, since the pair $(\Theta_{11},\left[\begin{array}{cc}R_{11}&0\end{array}\right])$ is controllable, the condition (\ref{TF_cond}) implies that the matrix $\tilde C_p = C_pP$ must be of the form
\begin{equation}
\label{Cpform}
\tilde C_p=\left[\begin{array}{cc}0&\tilde C_{p2}\end{array}\right].
\end{equation}
where the matrix $\tilde C_{p2}\in \rbb^{m \times n_{p2}}$ is of rank $m$. 
From this, it follows that the condition (\ref{CpJnull}) reduces to the condition
\begin{equation}
\label{Cp2tJnorm}
\tilde C_{p2} \Theta_{22} \tilde C_{p2}^T = 0.
\end{equation}
Now since $\Theta_{22}$ is nonsingular and $\tilde C_{p2}$ is of rank $m$, it follows that the matrix $\tilde C_{p2} \Theta_{22}$ is of rank $m$ and its null space is of dimension $n_{p2} -m$. However, since $\tilde C_{p2}$ is of rank $m$, the equation (\ref{Cp2tJnorm}) implies we must have $m \leq n_{p2} -m$ and hence we will require 
\[
m \leq \frac{n_{p2}}{2} = \frac{n_p - \mbox{rank} \mathcal{C}_r}{2}
\]
in order for the conditions (\ref{TF_cond}), (\ref{CpJnull}), (\ref{Cprank}) to be satisfied.

We now introduce a change of variables 
$$
\tilde x_p = P^Tx_p = \left[\begin{array}{c} \tilde x_{p1}\\ \tilde x_{p2}\end{array}\right]
$$
 to the system (\ref{plant}). It follows that
\begin{eqnarray}
\label{plant1}
\dot{\tilde x}_p =  \left[\begin{array}{c} \dot{\tilde x}_{p1}\\ \dot{\tilde x}_{p2}\end{array}\right]&=& P^T A_p P\tilde x_p \nonumber \\
&=& 2 P^T\Theta_p R_pP\tilde x_p = 2 P^T\Theta_p P P^T R_pP\tilde x_p\nonumber \\
&=& 2 \left[\begin{array}{cc} \Theta_{11} & 0 \\ 0 & \Theta_{22} \end{array}\right]\left[\begin{array}{cc}R_{p11}  & 0\\0 & 0\end{array}\right]\tilde x_p\nonumber \\
&=& \left[\begin{array}{c}2\Theta_{11}R_{p11} \tilde x_{p1}\\0\end{array}\right].
\end{eqnarray}
Also, 
\[
z_p = C_pP\tilde x_p = \left[\begin{array}{cc}0&\tilde C_{p2}\end{array}\right]\left[\begin{array}{c} \tilde x_{p1}\\ \tilde x_{p2}\end{array}\right] 
= \tilde C_{p2}\tilde x_{p2}
\]
using (\ref{Cpform}). 

It follows from (\ref{plant1}) that the plant variables $\tilde x_{p2}$ will remain constant while the variables $\tilde x_{p1}$ evolve dynamically for the plant system. Also, we have shown that the variables $z_p$ to be estimated  must be chosen to depend only on the variables $\tilde x_{p2}$ and not the variables $\tilde x_{p2}$. This will mean that if the quantum plant is a closed quantum system and not coupled to the quantum observer, the variables $z_p$ will remain constant. However, if the quantum plant is coupled to a quantum observer, this may longer apply. In the sequel, we will show that for a suitably designed quantum observer, the variables $z_p$ will remain constant even when the quantum plant is coupled to the quantum observer. 

\section{ Direct Coupling Coherent Quantum Observers}

We consider a {\em reduced order direct coupled linear quantum observer} defined by a symmetric matrix $R_o \in \rbb^{n_o
\times n_o}$, and matrices $R_c \in \rbb^{n_p \times n_o}$, $C_o\in \rbb^{m_p \times n_o}$. These matrices define an observer Hamiltonian
\begin{equation}
\label{observer_hamiltonian}
\mathcal{H}_o=\half x_o\trp R_o x_o,
\end{equation}
and a coupling Hamiltonian
\begin{equation}
\label{coupling_hamiltonian}
\mathcal{H}_c=\half x_p\trp R_c x_o+\half x_o\trp R_c\trp x_p.
\end{equation}
The matrix $C_o$ also defines the vector of output variables for the observer as $z_o(t)  = C_ox_o(t)$. 

The augmented quantum linear system consisting of the quantum plant and the direct coupled  quantum observer is then a quantum system of the form (\ref{quantum_system}) described by the total Hamiltonian
\begin{eqnarray}
\mathcal{H}_a &=& \mathcal{H}_p+\mathcal{H}_c+\mathcal{H}_o\nonumber \\
 &=& \half x_a\trp R_a x_a
\label{total_hamiltonian}
\end{eqnarray}
where
$x_a = \left[\begin{array}{l}x_p\\x_o\end{array}\right]$ and 
$R_a = \left[\begin{array}{ll}R_p & R_c\\R_c^T & R_o\end{array}\right]$. Then, using (\ref{eq_coef_cond_A}), it follows that the augmented quantum linear system is described by the equations
\begin{eqnarray}
\left[\begin{array}{l}\dot x_p(t)\\\dot x_o(t)\end{array}\right] &=& 
A_a\left[\begin{array}{l} x_p(t)\\ x_o(t)\end{array}\right];~ x_p(0)=x_{0p};~ x_o(0)=x_{0o};\nonumber \\
z_p(t) &=& C_px_p(t);\nonumber \\
z_o(t) &=& C_ox_o(t)
\label{augmented_system}
\end{eqnarray}
where $A_a = 2\Theta_a R_a$. Here 
\[
\Theta_a =\left[\begin{array}{ll}\Theta_p & 0\\0 & \Theta_o \end{array}\right].
\]

We now formally define the notion of a direct coupled linear quantum observer.

\begin{definition}
\label{D1}
The matrices $R_o \in \rbb^{n_o
\times n_o}$,  $R_c \in \rbb^{n_p \times n_o}$, $C_o\in \rbb^{m_p \times n_o}$ define a {\em direct coupled linear quantum observer} for the quantum linear plant (\ref{plant}) if the corresponding augmented linear quantum system (\ref{augmented_system}) is such that
\begin{equation}
\label{average_convergence}
\lim_{T \rightarrow \infty} \frac{1}{T}\int_{0}^{T}(z_p(t) - z_o(t))dt = 0.
\end{equation}
\end{definition}

\section{Constructing a Reduced Order Direct Coupling Coherent Quantum Observer}
\label{sec:construction}
In order to construct a reduced order direct coupled coherent observer, we assume that the quantum plant satisfies the conditions  (\ref{TF_cond}), (\ref{CpJnull}), (\ref{Cprank}) and apply the transformation $\left[\begin{array}{c} \tilde x_{p1}\\ \tilde x_{p2}\end{array}\right] = \tilde x_p = P^Tx_p$ considered in the previous section. Also, we assume that the coupling Hamiltonian $\mathcal{H}_c$ depends only on $\tilde x_{p2}$ and $x_o$ but not on $\tilde x_{p1}$; i.e., we can write
\begin{equation}
\label{coupling_hamiltonian1}
\mathcal{H}_c=\half \tilde x_{p2}\trp \tilde R_c x_o+\half x_o\trp \tilde R_c\trp \tilde x_{p2}
\end{equation}
where 
\begin{equation}
\label{Rc}
R_c = P \left[\begin{array}{l}0 \\ \tilde R_c\end{array}\right].
\end{equation}
Hence, we can write
\begin{eqnarray*}
\mathcal{H}_a &=& \half \tilde x_{p1}\trp R_{p11} \tilde x_{p1}+\half \tilde x_{p2}\trp \tilde R_c x_o+\half x_o\trp \tilde R_c\trp \tilde x_{p2} \nonumber \\
&&+ \half x_o\trp R_o x_o\nonumber \\
&=&\half \tilde x_a\trp \tilde R_a \tilde x_a
\end{eqnarray*}
where
$x_a = \left[\begin{array}{l}\tilde x_{p1}\\ \tilde x_{p2}\\x_o\end{array}\right]$ and 
$R_a = \left[\begin{array}{lll}R_{p11}&0 & 0\\0 & 0 & \tilde R_c \\
0 & \tilde R_c^T & R_o\end{array}\right]$.

We now suppose that 
\[
n_o = \left\{\begin{array}{l}m \mbox{ if $m$ is even;}\\m+1 \mbox{ if $m$ is odd.}\end{array}\right.
\]
Thus, $n_o$ is an even number and this corresponds to a reduced order quantum observer. 

We also suppose that the matrices $R_o$,  $\tilde R_c$, $C_o$ are such that 
\begin{equation}
\label{design1}
\tilde R_c =  \alpha\beta^T,~~ \alpha =  \tilde C_{p2}^T,~~ R_o > 0
\end{equation}
where $ \tilde C_{p2}^T \in \rbb^{n_{p2}\times m}$ and $\beta \in \rbb^{n_o \times m}$  is full rank. In addition, we write 
$\Theta =  \left[\begin{array}{lll}\Theta_{11} & 0 & 0\\0 & \Theta_{22} & 0 \\ 0 & 0 & \Theta_o\end{array}\right]$ where $\Theta_{11}$, $\Theta_{22}$ are defined as in (\ref{transformed_theta}) and 
$\Theta_o \in \rbb^{n_o\times n_o}$ is of the form (\ref{Theta}). Hence, the augmented system equations (\ref{augmented_system}) describing the combined plant-observer system imply
\begin{eqnarray}
\dot{\tilde x}_{p2}(t)&=& 2\Theta_{22} \alpha\beta^Tx_o(t);\nonumber \\
\dot x_o(t)&=&2 \Theta_o \beta\alpha^T\tilde{x}_{p2}(t)+2 \Theta_oR_ox_o(t);\nonumber \\
z_p(t) &=& \tilde C_{p2}\tilde x_p(t);\nonumber \\
z_o(t) &=& C_ox_o(t). 
\label{augmented_system1}
\end{eqnarray}

We will show that the given assumptions imply that the quantity $z_p(t) = \tilde C_{p2}\tilde x_{p2}(t)$ will be constant for the augmented quantum system (\ref{augmented_system1}). Indeed, it follows from (\ref{augmented_system1})  that
\[
\dot z_p(t) = 2\tilde C_{p2}\Theta_{22} \alpha\beta^Tx_o(t) = 2\tilde C_{p2}\Theta_{22}\tilde C_{p2}^T\beta^Tx_o(t) = 0
\]
using (\ref{Cp2tJnorm}). Therefore, 
\begin{equation}
\label{zp_const}
z_p(t) = z_p(0) = z_p
\end{equation}
for all $t\geq 0$. 

It now follows from (\ref{augmented_system1}) that 
\begin{eqnarray}
\label{xo1}
\dot x_o(t)&=&2 \Theta_o \beta \tilde C_{p2} \tilde{x}_{p2}(t)+2 \Theta_oR_ox_o(t)\nonumber \\
&=& 2 \Theta_oR_ox_o(t)+2 \Theta_o \beta z_p.
\end{eqnarray}
From this equation, we define the ``steady state'' value of the vector $x_o$ as
\[
\bar x_o = -R_o^{-1} \beta z_p.
\]
Then we define the ``error vector''
\[
\tilde x_o(t) = x_o(t) - \bar x_o.
\]
It follows from (\ref{xo1}) that $\tilde x_o(t)$ satisfies the differential equation
\begin{eqnarray*}
\dot{\tilde x}_o(t)&=& 2 \Theta_oR_ox_o(t)+2 \Theta_o \beta z_p \nonumber \\
&=& 2 \Theta_oR_o\tilde x_o(t)+ 2 \Theta_oR_o\bar x_o + 2 \Theta_o \beta z_p  \nonumber \\
&=& 2 \Theta_oR_o\tilde x_o(t).
\end{eqnarray*}

 We now show that 
\begin{equation}
\label{xtildeav}
\lim_{T \rightarrow \infty} \frac{1}{T}\int_{0}^{T}\tilde x_o(t)dt = 0
\end{equation}
following the proof of  a similar fact in \cite{PET14Ca}.  First note that the quantity $ \mathcal{\tilde H}_o(t) = \half \tilde x_o(t)\trp R_o \tilde x_o(t)$
remains constant in time. Indeed,
\begin{eqnarray*}
\frac{d}{dt}\mathcal{ \tilde H}_o(t) &=& \frac{1}{2}\dot{\tilde x}_o^TR_o\tilde x_o+\frac{1}{2}\tilde x_o^TR_o\dot{\tilde x}_o \nonumber \\
&=& -\tilde x_o^TR_o\Theta_o R_o \tilde x_o + \tilde x_o^TR_o\Theta_o R_o \tilde x_o = 0
\end{eqnarray*}
since $R_o$ is symmetric and $\Theta_o$ is skew-symmetric.
That is 
\begin{equation}
\label{Roconst}
\half \tilde x_o(t) \trp R_o \tilde x_o(t) = \half \tilde x_o(0) \trp R_o \tilde x_o(0) \quad \forall t \geq 0.
\end{equation}

However, $\tilde x_o(t) = e^{2\Theta_o R_ot}\tilde x_o(0)$ and $R_o > 0$. Therefore, it follows from (\ref{Roconst}) that
\[
\sqrt{\lambda_{min}(R_o)}\|e^{2\Theta_o R_ot}\tilde x_o(0)\| \leq \sqrt{\lambda_{max}(R_o)}\|\tilde x_o(0)\|
\]
for all $\tilde x_o(0)$ and $t \geq 0$. Hence, 
\begin{equation}
\label{exp_bound}
\|e^{2\Theta_o R_ot}\| \leq \sqrt{\frac{\lambda_{max}(R_o)}{\lambda_{min}(R_o)}}
\end{equation}
for all $t \geq 0$.

Now since $\Theta_o $ and $R_o$ are non-singular,
\[
\int_0^Te^{2\Theta_o R_ot}dt = \half e^{2\Theta_o R_oT}R_o^{-1}\Theta_o ^{-1} - \half R_o^{-1}\Theta_o ^{-1}
\]
and therefore, it follows from (\ref{exp_bound}) that
\begin{eqnarray*}
\lefteqn{\frac{1}{T} \|\int_0^Te^{2\Theta_o R_ot}dt\|}\nonumber \\
 &=& \frac{1}{T} \|\frac{1}{2}e^{2\Theta_o R_oT}R_o^{-1}\Theta_o ^{-1} - \frac{1}{2}R_o^{-1}\Theta_o ^{-1}\|\nonumber \\
&\leq& \frac{1}{2T}\|e^{2\Theta_o R_oT}\|\|R_o^{-1}\Theta_o ^{-1}\| \nonumber \\
&&+ \frac{1}{2T}\|R_o^{-1}\Theta_o ^{-1}\|\nonumber \\
&\leq&\frac{1}{2T}\sqrt{\frac{\lambda_{max}(R_o)}{\lambda_{min}(R_o)}}\|R_o^{-1}\Theta_o ^{-1}\|\nonumber \\
&&+\frac{1}{2T}\|R_o^{-1}\Theta_o ^{-1}\|\nonumber \\
&\rightarrow & 0 
\end{eqnarray*}
as $T \rightarrow \infty$. Hence,  
\begin{eqnarray*}
\lefteqn{\lim_{T \rightarrow \infty} \frac{1}{T}\|\int_{0}^{T} \tilde x_o(t)dt\| }\nonumber \\
&=& \lim_{T \rightarrow \infty}\frac{1}{T}\|\int_{0}^{T} e^{2\Theta_o R_ot}\tilde x_o(0)dt\| \nonumber \\
&\leq& \lim_{T \rightarrow \infty}\frac{1}{T} \|\int_{0}^{T} e^{2\Theta_o R_ot}dt\|\|\tilde x_o(0)\|\nonumber \\
&=& 0.
\end{eqnarray*}
This implies
\[
\lim_{T \rightarrow \infty} \frac{1}{T}\int_{0}^{T} \tilde x_o(t)dt = 0.
\]

Now we have 
\begin{eqnarray*}
\lim_{T \rightarrow \infty} \frac{1}{T}\int_{0}^{T} z_o(t)dt &=& \lim_{T \rightarrow \infty} \frac{1}{T}\int_{0}^{T} C_ox_o(t)dt\nonumber \\
&=& \lim_{T \rightarrow \infty} \frac{1}{T}\int_{0}^{T} C_o(\tilde x_o(t)+ \bar x_o)dt\nonumber \\
&=& \lim_{T \rightarrow \infty} \frac{1}{T}\int_{0}^{T} C_o \bar x_odt\nonumber \\
&=& C_o \bar x_o = -C_o R_o^{-1} \beta z_p.
\end{eqnarray*}
We now choose the matrices $C_o \in \rbb^{m\times n_o}$ and $\beta \in \rbb^{n_o \times m}$ so that
\begin{equation}
\label{design2}
-C_o R_o^{-1} \beta = I.
\end{equation}
This is always possible since $n_o \geq m$. It follows that 
\[
\lim_{T \rightarrow \infty} \frac{1}{T}\int_{0}^{T} z_o(t)dt = z_p
\]
and hence, the condition (\ref{average_convergence}) is satisfied. Thus, we have proved the following theorem.

\begin{theorem}
\label{T1}
Consider a quantum plant of the form (\ref{plant}) satisfying the conditions (\ref{TF_cond}), (\ref{CpJnull}), (\ref{Cprank}). Then  the matrices $R_o$, $\tilde R_c$, $C_o$ constructed as  in (\ref{design1}), (\ref{design2})
  will define a reduced order direct coupled quantum observer achieving time-averaged consensus convergence for this quantum plant.
\end{theorem}
\section{Illustrative Example}
We now present some numerical simulations to illustrate the reduced order direct coupled  quantum observer described in the previous section. We choose the quantum plant to have 
\[
R_p = \left[\begin{array}{llllll} 
1 & 1 & 1 & 1 & 1 & 1\\
1 & 1 & 1 & 1 & 1 & 1\\
1 & 1 & 1 & 1 & 1 & 1\\
1 & 1 & 1 & 1 & 1 & 1\\
1 & 1 & 1 & 1 & 1 & 1\\
1 & 1 & 1 & 1 & 1 & 1
\end{array}\right].
\]
Then, the corresponding matrix $\mathcal{C}_r$ defined in (\ref{scriptCr}) is given by
{\tiny
\[
\mathcal{C}_r = \left[\begin{array}{rrrrrrrrrrrr} 
1 & 1 & 1 & 1 & 1 & 1 & 1 & 1 & 1 & 1 & 1 & 1 \\
1 & 1 & 1 & 1 & 1 & 1 & -1 & -1 & -1 & -1 & -1 & -1\\
1 & 1 & 1 & 1 & 1 & 1 & 1 & 1 & 1 & 1 & 1 & 1 \\
1 & 1 & 1 & 1 & 1 & 1 & -1 & -1 & -1 & -1 & -1 & -1\\
1 & 1 & 1 & 1 & 1 & 1 & 1 & 1 & 1 & 1 & 1 & 1 \\
1 & 1 & 1 & 1 & 1 & 1 & -1 & -1 & -1 & -1 & -1 & -1
\end{array}\right].
\]
}
This matrix has rank $2$. From this, the orthogonal matrix $P$ is calculated by finding the svd of $\mathcal{C}_r$. This yields
{\tiny
\[
P = \left[\begin{array}{rrrrrr} 
   -0.5774 &  0.0000 &    0.5825  &  -0.5722  &   0.0000 &   -0.00001\\
         0 & -0.5774 &    0.5722  &   0.5825  &  -0.0000 &    0.00001\\
   -0.5774 & -0.0000 &   -0.2912  &   0.2861  &   0.6938 &   -0.13651\\
         0 & -0.5774 &   -0.2861  &  -0.2912  &  -0.1365 &   -0.69381\\
   -0.5774 & -0.0000 &   -0.2912  &   0.2861  &  -0.6938 &    0.13651\\
         0 & -0.5774 &   -0.2861  &  -0.2912  &   0.1365 &    0.6938
\end{array}\right].
\]
}
The corresponding transformed plant Hamiltonian matrix $\tilde R_p = P^TR_pP$ is in the form (\ref{Rpt}) where
\[
 R_{p11} = \left[\begin{array}{rr}
3 & 3 \\
 3 & 3
\end{array}\right].
\]
Also, the transformed commutation matrix $\tilde \Theta_p = P^T\Theta_p P$  is in the form (\ref{transformed_theta}) where
\[
\Theta_{11} = \left[\begin{array}{rr}
0 & 1\\-1 & 0
\end{array}\right], ~~ \Theta_{22} = \left[\begin{array}{rrrr}
0 & 1 & 0 & 0\\
-1 & 0 & 0 & 0 \\
0 & 0 & 0 & -1\\
0 & 0 & 1 & 0
\end{array}\right].
\]

In order to choose a suitable value of the matrix $C_p$ so that condition (\ref{TF_cond}) is satisfied, we choose $\tilde C_p = C_pP$  of the form (\ref{Cpform}) where $\tilde C_{p2} \in \rbb^{2 \times 4}$. Also, we require that the condition  (\ref{Cp2tJnorm}) is satisfied. It is straightforward to verify that this condition is satisfied by the matrix 
\[
\tilde C_{p2} = \left[\begin{array}{rrrr}
1 & 1 & 1 & 1\\
1 & 1 & -1 & -1
\end{array}\right].
\]
This corresponds to the matrix 
{\tiny
\[
C_p = \left[\begin{array}{rrrrrr}
    0.0103 &   1.1547 &    0.5522 &   -1.4076 &   -0.5625 &    0.2530\\
    0.0103 &   1.1547 &   -0.5625 &    0.2530 &    0.5522 &   -1.4076
\end{array}\right]
\]
}
which is such that conditions (\ref{TF_cond}), (\ref{CpJnull}), (\ref{Cprank}) are satisfied. 

The quantum plant defined by the matrices $R_p$ and $C_p$ given above is a plant of the form considered in Section \ref{sec:analysis} where $n_p = 6$, $n_{p1} = 2$, $n_{p2} = 4$, and $m= 2$. Hence, we will construct a reduced order observer as described in Section \ref{sec:construction} with $n_o = 2$. In order to construct the observer, we need to choose matrices $R_o > 0$, $\beta$ and $C_o$ such that (\ref{design2}) is satisfied. In this example, we will choose
\[
R_o = I,~~C_o = I,~~\beta=-I.
\]
Then the matrix $\tilde R_c$ is constructed according to (\ref{design1}) as
\[
\tilde R_c =\left[\begin{array}{rr}
    -1  &  -1\\
    -1  &  -1\\
    -1  &   1\\
    -1  &   1
\end{array}\right].
\]
From this, the matrix $R_c$ is constructed according to (\ref{Rc}) as
\[
R_c = \left[\begin{array}{rr}
   -0.0103 &  -0.0103\\
   -1.1547 &  -1.1547\\
   -0.5522 &   0.5625\\
    1.4076 &  -0.2530\\
    0.5625 &  -0.5522\\
   -0.2530 &   1.4076
\end{array}\right].
\]

 The augmented plant-observer system is described by the equations (\ref{augmented_system}). To simulate these equations we can write
\[
x_a(t)
= \Phi(t) x_a(0)
\]
where $\Phi(t) = e^{2\Theta_a R_a t}$. Furthermore, the plant variables to be estimated are given by
\[
z_p(t) = \left[\begin{array}{ll}C_p & 0 \end{array}\right] \Phi(t) x_a(0)
\]
and the observer output variables are given by
\[
z_o(t)  = \left[\begin{array}{ll}0 & C_o \end{array}\right] \Phi(t) x_a(0).
\]
Although the quantities $z_p(t)$ and $z_o(t)$ are operators which cannot be plotted directly, we can plot the coefficients in the above equations which define the components of $z_p(t)$ or  $z_o(t)$ with respect to the initial condition operators in $x_a(0)$. 

\begin{figure}[htbp]
\begin{center}
\includegraphics[width=7cm]{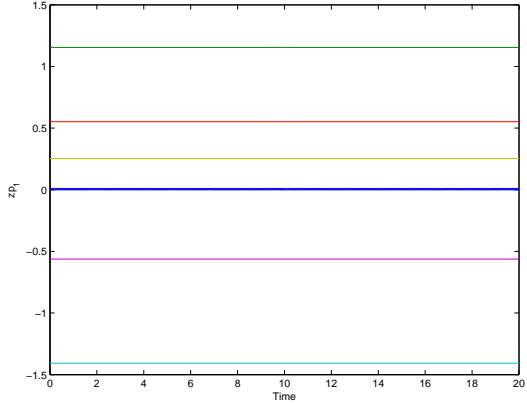}
\end{center}
\caption{Coefficient functions defining the first component of $z_p(t)$.}
\label{F1}
\end{figure}

\begin{figure}[htbp]
\begin{center}
\includegraphics[width=7cm]{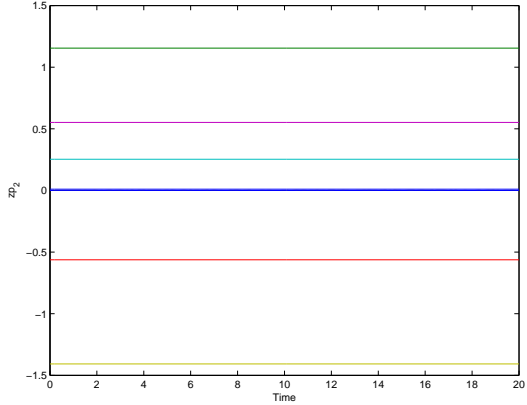}
\end{center}
\caption{Coefficient functions defining the second component of $z_p(t)$.}
\label{F2}
\end{figure}

In Figure \ref{F1}, we plot these coefficients corresponding to the first plant variable to be estimated.
In Figure \ref{F2}, we plot these coefficients corresponding to the second plant variable to be estimated.
These figures verify that the quantity $z_p(t)$ remains constant at its initial value. 

\begin{figure}[htbp]
\begin{center}
\includegraphics[width=7cm]{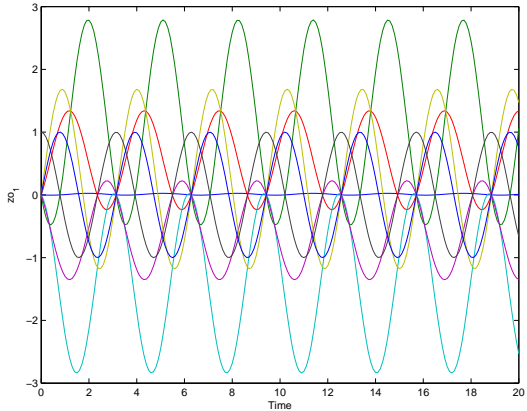}
\end{center}
\caption{Coefficient functions defining the first component of $z_o(t)$.}
\label{F3}
\end{figure}

\begin{figure}[htbp]
\begin{center}
\includegraphics[width=7cm]{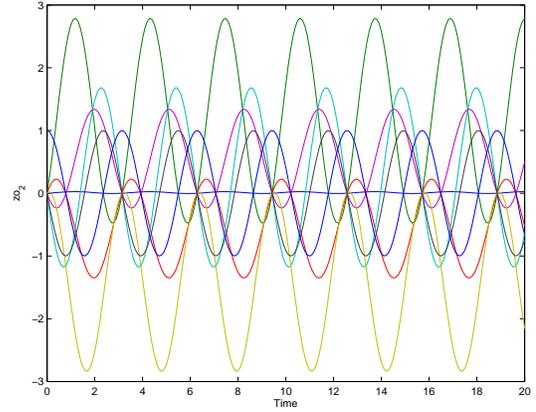}
\end{center}
\caption{Coefficient functions defining the second component of $z_o(t)$.}
\label{F4}
\end{figure}

In Figure \ref{F3}, we plot these coefficients corresponding to the first observer output variable, which is designed to provide an  estimate of the first plant variable to be estimated.
In Figure \ref{F4}, we plot these coefficients corresponding to the second observer output variable, which is designed to provide an  estimate of the second plant variable to be estimated.
From Figures \ref{F3} and \ref{F4}, we can see that $z_o(t)$  evolves in a time-varying and oscillatory way. 

\begin{figure}[htbp]
\begin{center}
\includegraphics[width=7cm]{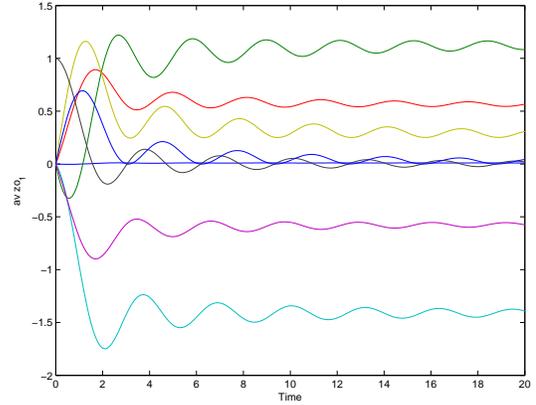}
\end{center}
\caption{Time averaged coefficient functions defining the first component of $z_o(t)$.}
\label{F5}
\end{figure}

To illustrate the time average convergence property of the quantum observer (\ref{average_convergence}), we now plot the time averaged quantities corresponding to Figures \ref{F3} and \ref{F4}. In Figure \ref{F5}, we plot the time averaged coefficients corresponding to the first observer output variable.
Comparing this figure with Figure \ref{F1}, we can see that the time average of the first component of $z_o(t)$ converges to the first component of $z_p$. 

\begin{figure}[htbp]
\begin{center}
\includegraphics[width=7cm]{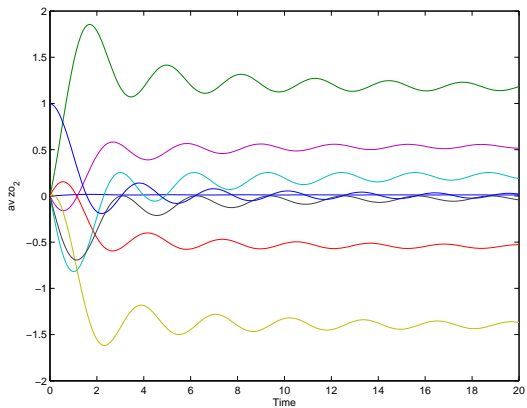}
\end{center}
\caption{Time averaged coefficient functions defining the second component of $z_o(t)$.}
\label{F6}
\end{figure}

In Figure \ref{F6}, we plot the time averaged coefficients corresponding to the second observer output variable.
Comparing this figure with Figure \ref{F2}, we can see that the time average of the second component of $z_o(t)$ converges to the second component of $z_p$.


\begin{thebibliography}{10}
\providecommand{\url}[1]{#1}
\csname url@rmstyle\endcsname
\providecommand{\newblock}{\relax}
\providecommand{\bibinfo}[2]{#2}
\providecommand\BIBentrySTDinterwordspacing{\spaceskip=0pt\relax}
\providecommand\BIBentryALTinterwordstretchfactor{4}
\providecommand\BIBentryALTinterwordspacing{\spaceskip=\fontdimen2\font plus
\BIBentryALTinterwordstretchfactor\fontdimen3\font minus
  \fontdimen4\font\relax}
\providecommand\BIBforeignlanguage[2]{{%
\expandafter\ifx\csname l@#1\endcsname\relax
\typeout{** WARNING: IEEEtran.bst: No hyphenation pattern has been}%
\typeout{** loaded for the language `#1'. Using the pattern for}%
\typeout{** the default language instead.}%
\else
\language=\csname l@#1\endcsname
\fi
#2}}

\bibitem{MJ12a}
Z.~Miao and M.~R. James, ``Quantum observer for linear quantum stochastic
  systems,'' in \emph{Proceedings of the 51st IEEE Conference on Decision and
  Control}, Maui, December 2012.

\bibitem{VP9a}
I.~Vladimirov and I.~R. Petersen, ``Coherent quantum filtering for physically
  realizable linear quantum plants,'' in \emph{Proceedings of the 2013 European
  Control Conference}, Zurich, Switzerland, July 2013, arXiv:1301.3154.

\bibitem{EMPUJ6a}
Z.~Miao, L.~A.~D. Espinosa, I.~R. Petersen, V.~Ugrinovskii, and M.~R. James,
  ``Coherent quantum observers for finite level quantum systems,'' in
  \emph{Australian Control Conference}, Perth, Australia, November 2013.

\bibitem{PET14Aa}
I.~R. Petersen, ``A direct coupling coherent quantum observer,'' in
  \emph{Proceedings of the 2014 IEEE Multi-conference on Systems and Control},
  Antibes, France, October 2014, also available arXiv 1408.0399.

\bibitem{PET14Ba}
------, ``A direct coupling coherent quantum observer for a single qubit finite
  level quantum system,'' in \emph{Proceedings of 2014 Australian Control
  Conference}, Canberra, Australia, November 2014, also arXiv 1409.2594.

\bibitem{PET14Ca}
------, ``Time averaged consensus in a direct coupled distributed coherent
  quantum observer,'' in \emph{Proceedings of the 2015 American Control
  Conference}, Chicago, IL, July 2015.

\bibitem{PET14Da}
------, ``Time averaged consensus in a direct coupled coherent quantum observer
  network for a single qubit finite level quantum system,'' in
  \emph{Proceedings of the 10th ASIAN CONTROL CONFERENCE 2015}, Kota Kinabalu,
  Malaysia, May 2015.

\bibitem{PeHun1a}
I.~R. Petersen and E.~H. Huntington, ``A possible implementation of a direct
  coupling coherent quantum observer,'' in \emph{Proceedings of 2015 Australian
  Control Conference}, Gold Coast, Australia, November 2015.

\bibitem{JNP1}
M.~R. James, H.~I. Nurdin, and I.~R. Petersen, ``${H}^\infty$ control of linear
  quantum stochastic systems,'' \emph{IEEE Transactions on Automatic Control},
  vol.~53, no.~8, pp. 1787--1803, 2008, arXiv:quant-ph/0703150.

\bibitem{GJ09}
J.~Gough and M.~R. James, ``The series product and its application to quantum
  feedforward and feedback networks,'' \emph{IEEE Transactions on Automatic
  Control}, vol.~54, no.~11, pp. 2530--2544, 2009.

\bibitem{ZJ11}
G.~Zhang and M.~James, ``Direct and indirect couplings in coherent feedback
  control of linear quantum systems,'' \emph{IEEE Transactions on Automatic
  Control}, vol.~56, no.~7, pp. 1535--1550, 2011.

\bibitem{RUG96}
W.~J. Rugh, \emph{Linear System Theory}, 2nd~ed.\hskip 1em plus 0.5em minus
  0.4em\relax Englewood Cliffs, N.J.: Prentice-Hall, 1996.

\end{thebibliography}

\end{document}